\documentclass[a4paper]{scrartcl}
\usepackage{array}
\usepackage{amsmath}
\usepackage[nolist]{acronym}
\usepackage[english]{babel}
\usepackage[margin=2cm]{geometry}
\usepackage{tikz}
\usepackage{pgfplots}

\title{Comparison between GFDM and VOFDM}
\author{Maximilian Matth\'e, Ivan Gaspar, Luciano Mendes, Dan
  Zhang\\\small Vodafone Chair Mobile Communication Systems, Technical
  University Dresden\\\small
  \texttt{firstname.lastname@ifn.et.tu-dresden.de}}
\date{}

\newcommand{\mat}[1]{\mathbf{#1}}
\def\F{\mat{F}}
\def\H{^H}
\def\I{\mat{I}}

\begin{document}
\maketitle

\begin{acronym}
  \acro{GFDM}{Generalized Frequency Division Multiplexing}
  \acro{VOFDM}{Vector-OFDM}
  \acro{VB}{vector blocks}
  \acro{SS}{subsymbol}
\end{acronym}

This document provides a comparison of the transmission techniques
used in \ac{GFDM} and \ac{VOFDM}.  Within the document both systems
are coarsely described and common and distinct properties are
highlighted.

\section{Mathematical Description of GFDM and VOFDM}
In the following we describe the signal processing necessary for
creating a transmit signal. For simplicity we only consider the pure
modulation, without addition of CP or more advanced waveform
engineering techniques.

\subsection{V-OFDM}
\ac{VOFDM} initially proposed in 2001 by X.G. Xia \cite{Xia2001} is
a multicarrier technique that is able to deal with
spectral nulls and better exploits frequency diversity than conventional OFDM.  Further studies on \ac{VOFDM} have been carried out in
e.g \cite{Cheng2011,Han2010,Li2012}. The following mathematical
description of the VOFDM transmitter is adapted from the derivation in
\cite{Li2012}.  However, in the present description we have focused on
the formulation of the transmitter using a single matrix
representation, which eases understanding of the system.

Consider a data vector $\vec{d}$ of length $N$,
\begin{align}
  \vec{d}&=[d_{0}, d_{1}, d_{2}, \dots d_{N-1}].
\end{align}
Now, this data vector is reshaped into a matrix $\mat{D}$ of $M$ rows and $L$
columns such that $N=LM$ and
\begin{align}
  \mat{D} &=
  \begin{pmatrix}
    d_{0} & d_{M} & d_{2M}&  \dots & d_{(L-1)M}\\
    d_{1} & d_{M+1} & d_{2M+1}& \dots & d_{(L-1)M+1} \\
    \vdots &\vdots & \vdots&\ddots& \vdots \\
    d_{M-1} & d_{2M-1} &d_{3M-1}& \dots  & d_{LM-1}
  \end{pmatrix}.
\end{align}
The columns of $\mat{D}$ denote the \ac{VB} of length $M$ and $L$
\acp{VB} form a \ac{VOFDM} frame. Then, at the transmitter an IFFT
along the rows of $\mat{D}$ (i.e. of the \acp{VB}) is performed and
the resulting time-domain signal $\vec{x}$ is given by stacking the
transformed \acp{VB} on top of each other.  This operation can be
written by
\begin{align}
  \vec{x} &= \text{vec}(\mat{D}\F_{L}\H),
\end{align}
where $\text{vec}$ performs the vectorization operation (i.e. stacking
the colums of the argument on top of each other) and $\F_{L}$ denotes
the L-point DFT matrix. Using properties of the Kronecker product $\otimes$, we
obtain the final description of the linear V-OFDM modulation
\begin{align}
  \vec{x} &= \underbrace{(\F_L\H\otimes \I_M)}_{{\mat{V}}}\vec{d},
\end{align}
and $\mat{V}$ denotes the VOFDM modulation matrix.
\subsection{GFDM}
GFDM was initially proposed in 2009 by G. Fettweis et. al
\cite{Fettweis2009a} as a flexible multicarrier technique for future wireless systems. Further research on GFDM has been published in e.g. \cite{Michailow2014,Matthe2014,Michailow2015} and others towards 5G systems. GFDM transmission can be described with several notations, such as
time-domain description as in \cite{Michailow2012e}, frequency-domain
description as in \cite{Michailow2012d} or a low-complexity
description in the time domain as in \cite{Gaspar2015}.  In the
following mathematical description we derive the GFDM modulation
matrix based on the description in \cite{Gaspar2015}.  There, the
transmit data $\vec{d}_{m}$ for each subsymbol is repeated over the
full block and then multiplied by the pulse shaping filter which is
shifted to the corresponding subsymbol position. Consider a GFDM data
block $\mat{D}$
\begin{align}
  \mat{D} &=
  \begin{pmatrix}
    \vec{d}_0 & \vec{d}_1 & \dots & \vec{d}_{M-1}
  \end{pmatrix}=
  \begin{pmatrix}
    d_{0,0} & d_{0,1} & \dots & d_{0,M-1}\\
    d_{1,0} & d_{1,1} & & \vdots \\
    \vdots& & \ddots \\
    d_{K-1,0} & d_{K-1,1} & \dots & d_{K-1,M-1}
  \end{pmatrix},
\end{align}
where $K$ denotes the number of subcarriers and $M$ denotes the number
of \acp{SS}. An K-point IFFT is performed along
the columns of $\mat{D}$, i.e. separately for each \ac{SS}.  Then each
\ac{SS} is repeated $M$ times.  Finally each \ac{SS} is multiplied
element-wise by a circularly rotated version of the pulse shaping
filter and all subsymbols are summed up to generate the GFDM transmit
signal.
\begin{align}
  \vec{x}&=\sum_{m=0}^{M-1}
  \text{diag}(\mat{C}_{mK}\vec{g})\mat{R}_M\F_K\H\vec{d}_m\\
   &=
   \underbrace{\begin{bmatrix}
     \text{diag}(\mat{C}_{0}\vec{g})& \text{diag}(\mat{C}_{K}\vec{g})& \dots &\text{diag}(\mat{C}_{(M-1)K}\vec{g})
   \end{bmatrix}}_{\mat{A}}\text{vec}(\mat{D}),
\end{align}
where $\mat{R}_{M}=\I_{K}\otimes \mat{1}_{M}$ performs a M-fold
repetition of its argument, $\mat{C}_l$ performs a circular shift of $l$ elements of its
argument and $\text{diag}(\cdot)$ returns a diagonal matrix with its
argument on the diagonal. $\mat{A}$ is the resulting GFDM modulation
matrix, that is also derived in \cite{Michailow2012e} and also used in \cite{Matthe14}.

\section{Interperation and Comparison of Both Systems}
V-OFDM modulation defines data in the frequency domain in the rows of
$\mat{D}$. The data is tranferred to time domain by IFFT operation on
the rows of $\mat{D}$. Finally, reading the resulting
matrix column-wise, the corresponding time-domain signals for each row
in $\mat{D}$ are transmitted in an interleaved way. This can be
understood as if each time-domain signal is upsampled by factor $L$ and
transmitted with a delay equal to its row number, resembling a time domain multiplexing (TDM) system.

\begin{figure}
  \centering
  \includegraphics{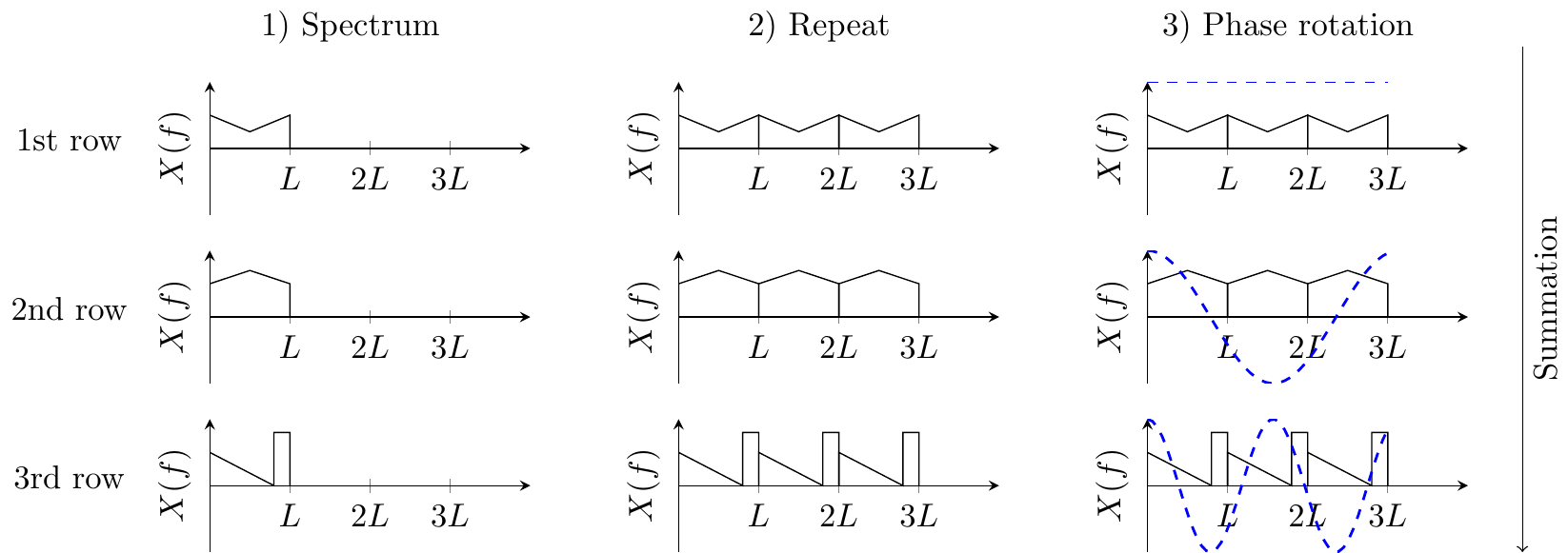}
  \caption{Illustration of VOFDM modulation in the frequency domain, for $M=3$.}
\label{fig:vofdm}
\end{figure}

The upsampling operation in time corresponds to a spectrum repetition.  Moreover,
transmission with a time offset corresponds to a phase rotation in the
frequency domain. Hence, the V-OFDM modulation can be illustrated as
in Fig. \ref{fig:vofdm}. For each row in $\mat{D}$, the spectrum is
repeated $M$ times and afterwards depending on the row index the phase
is rotated. Eventually, all spectra are summed up to yield the
complete transmit signal in the frequency domain.  From this
description we see that each data symbol in $\mat{D}$ is spread onto
the entire transmit spectrum, yielding frequency diversity in
frequency selective channels.  On the other hand, no spectral agility
is available, as always the full bandwith is occupied, regardless of
symbol allocation.

\begin{figure}[t]
  \centering
  \includegraphics{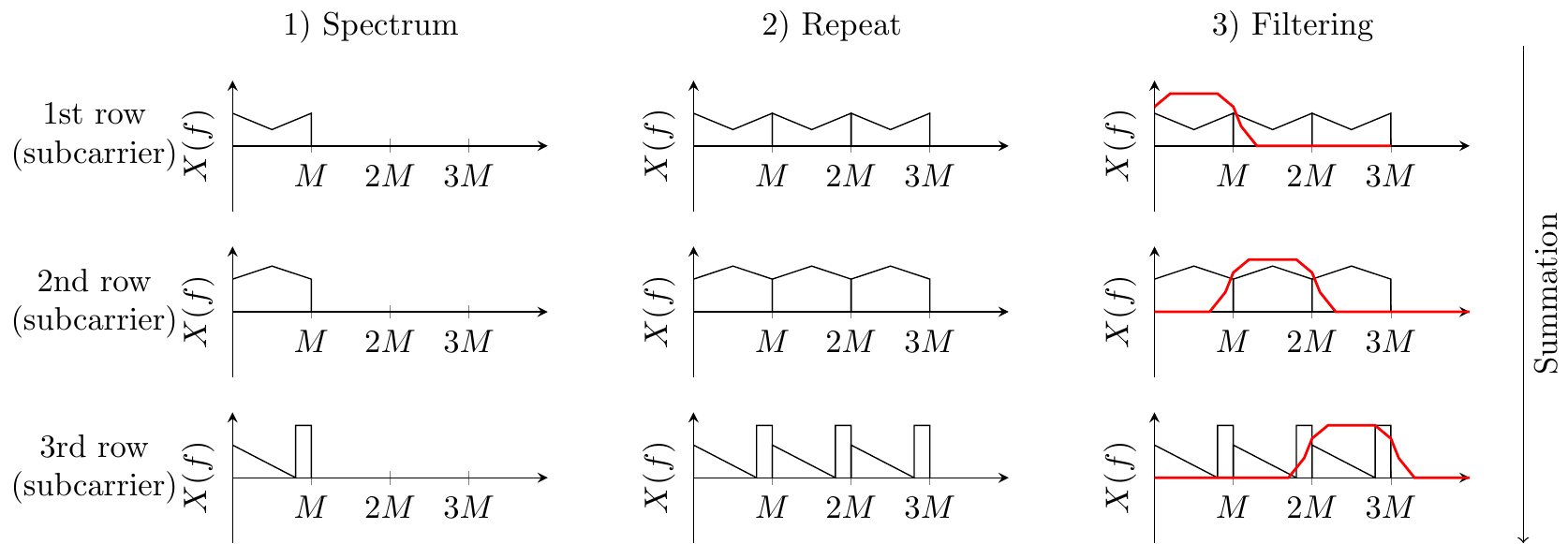}
  \caption{Illustration of GFDM modulation in the frequency domain, for $K=3$}
  \label{fig:gfdm}
\end{figure}

In contrast, the modulation of GFDM can also be understood as explicitly separating the signals in frequency domain in sub-bands. In GFDM the upsampled data on each subcarrier is
circularly convolved with the pulse shaping filter that is shifted to
the appropriate subcarrier frequency. The cyclic time convolution turns to
element-wise multiplication in the frequency domain. Furthermore,
upsampling in the time domain again corresponds to repetition in the
frequency domain as in the VOFDM system.  This technique is
illustrated in Fig. \ref{fig:gfdm}.

Hence, similar to both GFDM and VOFDM is a repetition of the
transmitted spectrum of one data matrix row \footnote{Ignoring the
  domain where the data is defined. For VOFDM its defined in the
  frequency domain, whereas in GFDM data on one subcarrier is defined
  in time domain}. The severe difference between both systems is the
applied window for element-wise multiplication in the frequency
domain, corresponding to convolution in time domain.

In GFDM, the signal is multiplied with a spectrally localized window,
that corresponds to a smooth function in the time domain (i.e. the
pulse shaping filter). In contrast, VOFDM utilizes a rotating
exponential that corresponds to a Dirac function in the time domain.
Accordingly, in GFDM both subsymbols and subcarriers are localized in
time and frequency domains, compared to VOFDM where all data spreads
over the entire spectrum and VB are interleaved in the time
domain. From the medium access point of view, VOFDM is more like CDMA using different spreading codes. GFDM is more like FDMA.

\begin{table}[t]
\caption{Comparison of GFDM and VOFDM.}
\label{tab:compare}  \centering
  \begin{tabular}[t]{p{4.7cm}|>{\centering\arraybackslash}m{5.5cm} >{\centering\arraybackslash}m{5.8cm}}
  Aspect &GFDM & VOFDM \\\hline
  Frequency domain structure & \multicolumn{2}{c}{Spectrum repetition for each data matrix row}\\\hline
  Frequency domain window & localized (sparse) filter & rectangular of full bandwidth\\\hline
  Filter change row to row&circular shift by one
  subcarrier&multiplication with complex exponential of increasing frequency\\\hline
  Localization & localized in time and frequency & interleaved in time and frequency\\\hline
\end{tabular}
\end{table}

Accordingly we can conclude that GFDM and VOFDM have similarities in
their spectral structure. However, the most significant difference is
the used frequency domain window.  The localized window of partially
GFDM wastes frequency diversity compared to VOFDM. On the other hand,
it allows to implement spectrally agile systems that leave certain
subcarriers empty for other systems. A comparison summary is given in
Tab. \ref{tab:compare}.

\bibliographystyle{plain}
\bibliography{library}

\end{document}